\documentclass[a4paper,12pt]{article}

\usepackage[a4paper,left=80pt,right=80pt,top=90pt,bottom=90pt]{geometry}
\usepackage[latin1]{inputenc}
\usepackage{dsfont}
\usepackage{amsfonts,amsmath,amssymb}
\usepackage{amsthm}
\usepackage{graphicx}
\usepackage[small,bf]{caption}
\usepackage{subcaption}
\usepackage{booktabs}
\usepackage[numbers, sort&compress]{natbib}
\allowdisplaybreaks[1]
\usepackage{color}
\usepackage{ulem}
\usepackage{tikz}
\usepackage{multirow}
\usetikzlibrary{arrows}
\usepackage{xspace}

\def\bal#1\eal{\begin{align}#1\end{align}}
\newcommand{\be}{\begin{equation}}
\newcommand{\ee}{\end{equation}}
\newcommand{\bea}{\begin{eqnarray}}
\newcommand{\eea}{\end{eqnarray}}
\newcommand{\besub}{\begin{subequations}}
\newcommand{\eesub}{\end{subequations}}
\newcommand{\ba}{\begin{array}}
\newcommand{\ea}{\end{array}}
\newcommand{\bi}{\begin{itemize}}
\newcommand{\ei}{\end{itemize}}
\newcommand{\nn}{\nonumber}

\newcommand{\Mcal}{{\cal M}}

\newcommand{\Lcal}{{\cal L}}

\newcommand{\lh}{\ensuremath{\lambda_H}}
\newcommand{\ls}{\ensuremath{\lambda_S}}
\newcommand{\lsp}{\ensuremath{\lambda^{\prime}_S}}
\newcommand{\lspp}{\ensuremath{\lambda^{\prime \prime}_S}}
\newcommand{\lhs}{\ensuremath{\lambda_{HS}}}
\newcommand{\lhsp}{\ensuremath{\lambda^{\prime}_{HS}}}

\begin{document}

\begin{titlepage}

\flushright{HIP-2017-20/TH\\
TUM-HEP/1091/17}

\vspace*{1cm}

\begin{center}
{\LARGE 
{\bf
A cancellation mechanism for

\vspace{6pt}
 dark matter--nucleon interaction
}
}
\\
[1.5cm]

{
{\bf
Christian Gross$^{1}$, Oleg Lebedev$^{1}$, Takashi Toma$^{2}$
}}
\end{center}

\vspace*{0.5cm}

\centering{
$^{1}$ ~\it{Department of Physics and Helsinki Institute of Physics, \\
Gustaf H\"allstr\"omin katu 2, FI-00014 Helsinki, Finland
}

$^{2}$~\it{
Physik-Department T30d, Technische Universit\"at M\"unchen,\\
 James-Franck-Stra\ss{}e, D-85748 Garching, Germany
}
}

\vspace*{1.6cm}

\begin{abstract}
\noindent
We consider a simple Higgs portal dark matter model, where the Standard Model is supplemented with a complex scalar whose imaginary part plays the role of WIMP dark matter (DM). We show that the direct DM detection cross section vanishes at tree level and zero momentum transfer due to a cancellation by virtue of a softly broken symmetry. This cancellation is operative for any mediator masses. As a result, our electroweak scale dark matter satisfies all of the phenomenological constraints quite naturally.
\end{abstract}

%\today

\end{titlepage}
\newpage

%\tableofcontents

%=========================================================================
%=========================================================================
\section{Introduction}
%=========================================================================
%=========================================================================

The ``Higgs portal''~\cite{Patt:2006fw} approach is a promising venue for addressing the problem of dark matter. It assumes that the only connection between the observable and dark sectors is
provided by the Higgs field. 
In this case, dark matter (DM) can belong to the Weakly Interacting Massive Particle (WIMP) category with the feature that the DM scattering on nucleons is suppressed compared to that of the standard WIMP due to the small Higgs--nucleon coupling. Also, the collider constraints on such models are rather weak since DM production is mediated by the Higgs field. This makes the Higgs portal an attractive framework which naturally satisfies many phenomenological constraints.

The simplest models include a real or complex singlet DM, see e.g.~\cite{Silveira:1985rk,McDonald:1993ex,Burgess:2000yq,Barger:2008jx,Djouadi:2011aa,Gabrielli:2013hma,Cline:2013gha,Khoze:2013uia}, where dark matter stability is due to $Z_2$ or global U(1) symmetries in the dark sector. 
 If the latter is endowed with gauge symmetry, vector Higgs portal dark matter arises naturally~\cite{Hambye:2008bq,Lebedev:2011iq,Gross:2015cwa}. In this case, the stabilizing symmetries are the discrete and continuous symmetries inherent in the Yang--Mills and U(1) systems. Finally, fermionic dark matter is also possible~\cite{Baek:2011aa,LopezHonorez:2012kv} with the relevant symmetry being the corresponding fermion number.

 Recently, the WIMP paradigm and Higgs portal dark matter, in particular, found themselves under pressure from ever--improving direct DM detection bounds~\cite{Aprile:2017iyp}. 
In the simplest models, the preferred DM mass range is pushed towards TeV values, although lower values cannot be excluded at the moment~\cite{Athron:2017kgt}. This raises the question whether 
there are classes of models where electroweak scale DM satisfies the direct detection constraints naturally. The answer to this question is affirmative. An example of such a class is provided by the ``secluded dark matter'' framework~\cite{Pospelov:2007mp} whose main feature is DM annihilation into unstable hidden sector states and which 
is natural in the Higgs portal construction~\cite{Arcadi:2016qoz}. 
Other possibilities explored in the literature include models with special parameter choices, for instance, in order to facilitate the co--annihilation processes~\cite{Casas:2017jjg} or take advantage of some cancellations in the direct detection
amplitude~\cite{Arcadi:2016kmk}.

In this work, we suggest a different possibility and present a very simple Higgs portal model where the direct DM detection amplitude is suppressed due to a cancellation by virtue of a softly broken symmetry.
The cancellation requires no tuning and takes place for any parameter choice.
As a result, electroweak scale WIMP dark matter is found to be consistent with all of the constraints, thereby underscoring the appeal of the WIMP paradigm.

%=========================================================================
%=========================================================================
\section{Higgs portal and a complex scalar}
%=========================================================================
%=========================================================================

Consider an extension of the Standard Model with a complex scalar $S$
interacting via the Higgs portal. Let us assume that the system is invariant under a global U(1) $S\rightarrow e^{i\alpha}S$, which is broken softly by a mass term for $S$:
\bal
V&=V_0 + V_{\rm soft} \;,\nonumber\\
V_0&= 
-\frac{ \mu_H^2}{2} \, |H|^2 
- \frac{\mu_S^2}{2} \, |S|^2 
+\frac{\lh}{2} |H|^4 
+\lhs |H|^2 |S|^2 
+ \frac{\ls}{2} |S|^4 \;, \nonumber\\
V_{\rm soft} &=
- \frac{\mu_S^{\prime 2}}{4} \, S^2 + \textrm{h.c.} \label{pot}
\eal
At the moment, we neglect higher dimension U(1) breaking operators which can be justified by treating the couplings as spurions (to be discussed in Section \ref{spurion}).
Also, we are assuming that the term linear in $S$ is forbidden by a $Z_2$ subgroup of the U(1), which remains unbroken in the spurion formalism. (The domain wall problem associated with the $Z_2$ breaking by $\langle S \rangle $ is avoided if U(1) is gauged in the UV--completion.)
 
 The parameter $\mu_S^{\prime 2}$ can always be made real and positive by phase redefinition. Thus, the system is invariant 
under the ``{\it CP}--symmetry''
\begin{equation}
S \to S^* \;.
\end{equation}
This symmetry remains unbroken by the $S$ vacuum expectation value
since for positive $\mu_S^{\prime 2}$ the vacuum expectation value (VEV) is $real$. 
It is due to the fact that only the $\mu_S^{\prime 2}$ term is sensitive to the phase of the $S$ field and the dependence is $-\cos( 2 \, {\rm Arg} S )$.
This immediately implies stability of the imaginary component of $S$
which plays the role of dark matter in our model.

Let us analyze the spectrum of the model.
Decomposing $S$ as
\be
S=(v_s+s+ i \chi)/\sqrt{2} \,,
\ee
with real $v_s$ and $\chi$ being dark matter, and 
$H^T=(0,v+h)/\sqrt{2}$, we find the following stationary point conditions at $h=0,s=0$:
\bal
\mu_H^2 &= \lh v^2 + \lhs v_s^2 \;,
\nn \\
\mu_S^2 &= \lhs v^2 + \ls v_s^2 - \mu_S^{\prime 2} \,.
\eal
Using these relations, the mass matrix for the {\it CP}-even states $(h,s)$ is found to be
\bal
\Mcal^2= \left( \begin{array}{cc}
\lh v^2 &\lhs v v_s \\
 \lhs v v_s & \ls v_s^2
\end{array} \right) \,,
\eal
while the mass of the pseudoscalar $\chi$ is
\be
m_\chi^2 = \mu_S^{\prime 2} \,.
\ee
$\Mcal^2$ can be diagonalised by the orthogonal transformation
$O^T \Mcal^2 O = \textrm{diag}(m_{h_1}^2,m_{h_2}^2) ,$
where
\bal
O&= \left( \begin{array}{cc}
\cos \theta & \sin \theta \\
- \sin \theta & \cos \theta
\end{array} \right) \;
\eal
and the angle $\theta$ satisfies
\bal
\tan 2 \theta &= \frac{2 \lhs v v_s }{\ls v_s^2 -\lh v^2} \,. \label{tantwotheta}
\eal
The mass squared eigenvalues are given by
\bal
m_{{h_1,h_2}}^2 =\frac12 \left( \lh v^2+\ls v_s^2\mp \frac{ \ls v_s^2 - \lh v^2}{\cos 2 \theta} \right) \,.
\label{evalues}
\eal
We identify $h_1$ with the 125~GeV Higgs boson. 
This leaves 4 free parameters: $m_{h_2}$, $m_\chi$, $\sin \theta$ and $v_s$.

%=========================================================================
%=========================================================================
\section{Cancellation in the direct detection amplitude}
%=========================================================================
%=========================================================================

\begin{figure}[h]
\centering{
\includegraphics[scale=0.5]{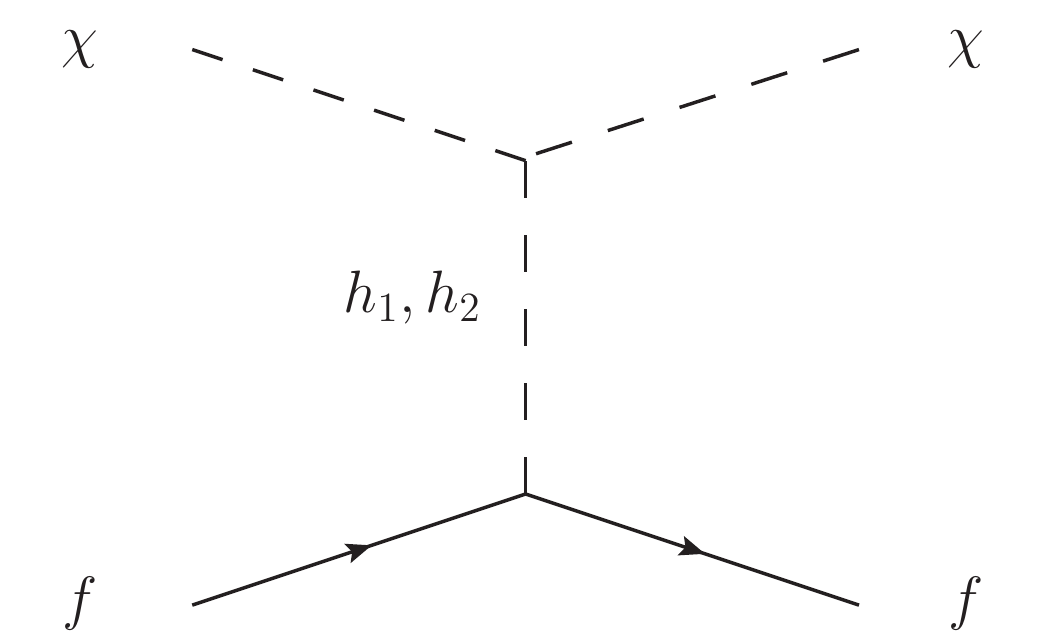}
 }
\caption{Tree--level dark matter scattering off SM matter.
}
\label{DD}
\end{figure}

The tree-level diagrams for scattering of $\chi$ on matter involve the $t$-channel exchange of a single $h_1$ or $h_2$ (Fig.~\ref{DD}).
The $\chi$-$\chi$-$h_{1,2}$ couplings are given by
\bea \label{kappa}
\mathcal{L}\supset \frac{v_s}{2} \chi^2 \left( \kappa_{\chi \chi h_1} \, h_1 + \kappa_{\chi \chi h_2} \, h_2 \right) \,,
\eea
with 
\bal
\kappa_{\chi \chi h_1}&= +\,\, m_{h_1}^2/v_s^2 \ \sin \theta \;,
\nn \\
\kappa_{\chi \chi h_2}&= - \, m_{h_2}^2/v_s^2 \ \cos \theta \,,
\eal 
whereas the couplings of $h_{1,2}$ to fermions $f$ are given by 
\be
\Lcal \supset - ( h_1 \cos \theta + h_2 \sin \theta) \sum_f \frac{m_f}{v} \bar f f \,.
\ee
Thus, the tree-level direct detection scattering amplitude is
\be
\mathcal{A}_{dd}(t) \propto \sin \theta \cos \theta \left(\frac{m_{h_2}^2}{t-m_{h_2}^2}-\frac{m_{h_1}^2}{t-m_{h_1}^2}\right) \simeq \sin \theta \cos \theta \ \frac{t \, (m_{h_2}^2-m_{h_1}^2)}{m_{h_1}^2 \ m_{h_2}^2} \simeq 0 
\ee
because the momentum transfer in this process is negligibly small, $t\simeq 0$. Thus, the contributions from the $h_1$-exchange and the $h_2$-exchange cancel each other up to tiny corrections of order $t/(100 \ {\rm GeV})^2$. Note that this does not require any relation between $m_{h_1}$ and $m_{h_2}$, and the cancellation occurs for $any$ choice of model parameters.

It is instructive to examine the cancellation mechanism 
 in the interaction basis, i.e. in terms of the states $h$ and $s$,
 where only $h$ couples to SM fermions.
 The relevant $\chi$-$\chi$-$h$ and $\chi$-$\chi$-$s$ couplings are
\bea
\mathcal{L}\supset -\frac{1}{2} \chi^2 \left(\lhs v \, h + \ls v_s \, s \right) \,,
\eea
while, for vanishing momentum transfer $t$, the propagator matrix is proportional to
\be
(\Mcal^2)^{-1}=\frac{1}{\det \Mcal^2} 
 \left( \begin{array}{cc}
\ls v_s^2 &-\lhs v v_s \\
- \lhs v v_s & \lh v^2
\end{array} \right) \,.
\ee
Since the SM fermions do not couple to $s$, the tree-level direct detection amplitude at $t=0$ indeed vanishes:
\be
\mathcal{A}_{dd} \propto 
 \left( \begin{array}{cc}
\lhs v , & \ls v_s 
\end{array} \right)
.
\left( \begin{array}{cc}
\ls v_s^2 &-\lhs v v_s \\
- \lhs v v_s & \lh v^2
\end{array} \right)
.
\left( \begin{array}{c}
1 \\
0 
\end{array} \right)
=0 \,.
\ee

The cancellation is due to the structure of the potential Eq.~$(\ref{pot})$ where the U(1) symmetry is broken only by the mass term.
This can be traced back to the \mbox{(pseudo-)} Goldstone nature of dark matter: $\chi$ is equivalent to the angular component of $S=\rho e^{i\phi}$, $\phi$, whose interactions vanish at zero momentum transfer. Introduction of the mass term $S^2$ does not affect the relevant vertex $\phi \phi \rho$, which vanishes for $\phi$ on-shell and zero momentum of $\rho$.
 U(1) breaking terms of higher dimension spoil the cancellation, however, as we show later, these can be highly suppressed when the 
 couplings are treated as spurions. We note that a (technically) similar cancellation was observed in~\cite{Arcadi:2016kmk}, although it occurred for a specific parameter choice and was not based on symmetry. 
 
 The cancellation is also spoiled by loop effects. In particular, higher dimension U(1) breaking terms are always generated at one loop. We discuss those in the next section.
There are also further 1--loop corrections, not related to U(1) breaking.
The largest of them only modify the $hNN$ vertex at zero momentum transfer and thus do not affect the cancellation. Other corrections involve multiple Higgs couplings to fermions and are subleading as long as $\lambda_S$ is relatively large.
A complete analysis of loop corrections is beyond the scope of this work and we restrict ourselves to the loop effects due to higher dimension U(1) breaking operators.

%=========================================================================
%=========================================================================
\section{Effect of higher dimension U(1) breaking terms}
%=========================================================================
%=========================================================================

At one loop, the following dimension--4 U(1) breaking terms are generated: 
\bal \label{delpot}
 V_{1}=
 \frac {\lhsp}2 |H|^2 S^2
+\frac{\lspp}{4} |S|^2 S^2
+\frac{\lsp}{4} S^4 
+ \textrm{h.c.} 
\eal
The couplings vanish in the U(1) symmetric limit $\mu_S^{\prime 2} \rightarrow 0$ and 
 are given by\footnote{This provides a good estimate of the loop effects for $\mu_S^{\prime 2}<\mu_S^{ 2}$. A more precise result can be obtained via the Coleman--Weinberg effective potential expansion around the true vacuum. 
The full analysis of loop corrections will be performed in our subsequent work.}
\bal
\lhsp &= { \lambda_{HS} \lambda_S \over 32 \pi^2} ~\ln { \mu_S^2 + \mu_S^{\prime 2}
\over \mu_S^2 - \mu_S^{\prime 2} } ~,\nonumber \\
\lspp &= { \lambda_S^2 \over 8 \pi^2} ~\ln { \mu_S^2 + \mu_S^{\prime 2}
\over \mu_S^2 - \mu_S^{\prime 2} } ~,\nonumber \\
\lsp &= { \lambda_S^2 \over 64 \pi^2} ~ \left( {\mu_S^2 \over \mu_S^{\prime 2}} 
 \ln { \mu_S^2 - \mu_S^{\prime 2}
\over \mu_S^2 + \mu_S^{\prime 2}} +2 
\right)~.
\eal
They are all real and do not spoil the symmetry $S \rightarrow S^*$, nor is this symmetry broken by the vacuum. 
Let us summarize the changes in the spectrum and couplings induced by $V_1$. The stationary point conditions at $h=0, s=0$ now become 
\bal
 \mu_H^2 &= \lh v^2 + (\lhs + \lhsp ) v_s^2 ~,
\nn \\
 \mu_S^2 &= (\lhs + \lhsp ) v^2 + (\ls + \lsp + \lspp ) v_s^2 -\mu_S^{\prime 2} \,,
\eal
while the $(h,s)$ mass matrix is 
\bal
\Mcal^2= \left( \begin{array}{cc}
\lh v^2 &\left(\lhs + \lhsp \right)v v_s \\
 \left(\lhs + \lhsp \right) v v_s & \left(\ls + \lsp + \lspp \right) v_s^2
\end{array} \right) \,.
\eal
The expressions for the mass squared eigenvalues as well as the mixing angle $\sin \theta$ are therefore obtained by replacing $\ls \to \ls + \lsp + \lspp $ and $\lhs \to \lhs + \lhsp $ in Eqs.~(\ref{tantwotheta},\ref{evalues}).
The dark matter mass becomes
\be
m_\chi^2 = \mu_S^{\prime 2} - \lhsp v^2 - \left(2 \lsp + \lspp /2 \right) v_s^2\,.
\ee
The most important effect of the new terms is that they modify the dark matter couplings to $h_{1,2}$ in Eq.~(\ref{kappa}):
\bal
\kappa_{\chi \chi h_1}&= 
+ \sin \theta 
\left(
\frac{m_{h_1}^2}{v_s^2}
-4 \lsp - \lspp
\right)
+ 
 \frac{2 \lhsp v }{v_s} \cos \theta ~,
\nn \\
\kappa_{\chi \chi h_2}&= 
- \cos \theta 
\hspace{-1.5pt}
\left(
\frac{m_{h_2}^2}{v_s^2} 
-4 \lsp - \lspp
\right)
+ 
 \frac{2 \lhsp v }{v_s} \sin \theta
~.
\eal 
Obviously, the extra terms in $\kappa_{\chi \chi h_{1,2}}$ do not cancel in the 
direct detection amplitude in general. However, this effect is loop suppressed resulting in very small DM detection rates, which we quantify in the next section.

%=========================================================================
%=========================================================================
\section{Parameter space analysis}
%=========================================================================
%=========================================================================

In this section, we perform a numerical analysis of the relevant constraints on the model, using the software {\it Micromegas}~\cite{Belanger:2014vza}.
Our dark matter candidate $\chi$ belongs to the WIMP category and 
we impose the PLANCK constraint $\Omega h^2 =
0.1197\pm 0.0022$~\cite{Ade:2015xua} at $3\sigma$ on its relic abundance. The most stringent direct DM detection bound is due to XENON1T~\cite{Aprile:2017iyp}. Also, one needs to make sure that the perturbative calculations can be trusted, which can be interpreted as the perturbative unitarity constraint 
$\lambda_S<8\pi/3$~\cite{Chen:2014ask} derived from $h_2h_2\to h_2h_2$ scattering at high energies. Finally, if dark matter is light, it can affect the LHC Higgs signal strength
$\mu=1.09^{+0.11}_{-0.10}$~\cite{Khachatryan:2016vau} via invisible Higgs decay.
This results in the bound $\mathrm{Br}(h_1\to\text{inv})\leq0.11$ at $95\%$ Confidence
Level.

\begin{figure}
\centering{
\includegraphics[width=0.495\textwidth]{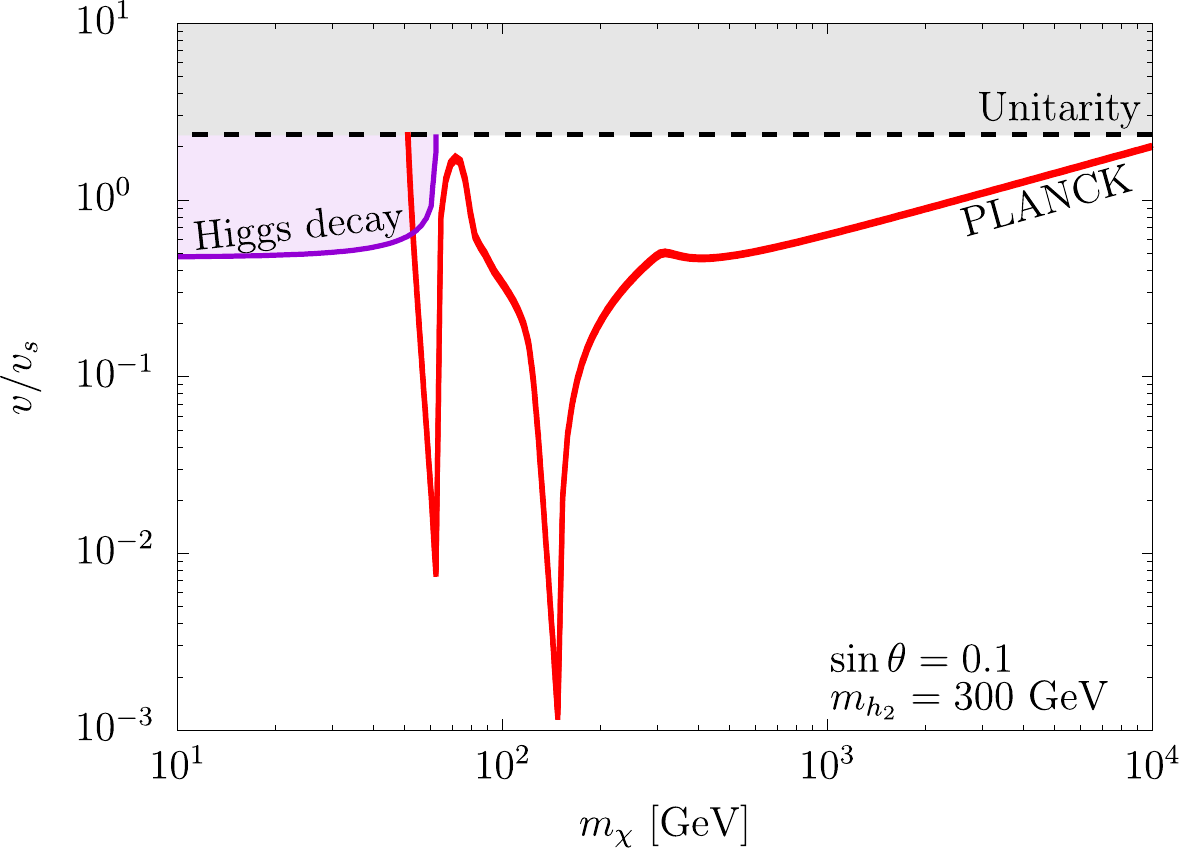}
\includegraphics[width=0.495\textwidth]{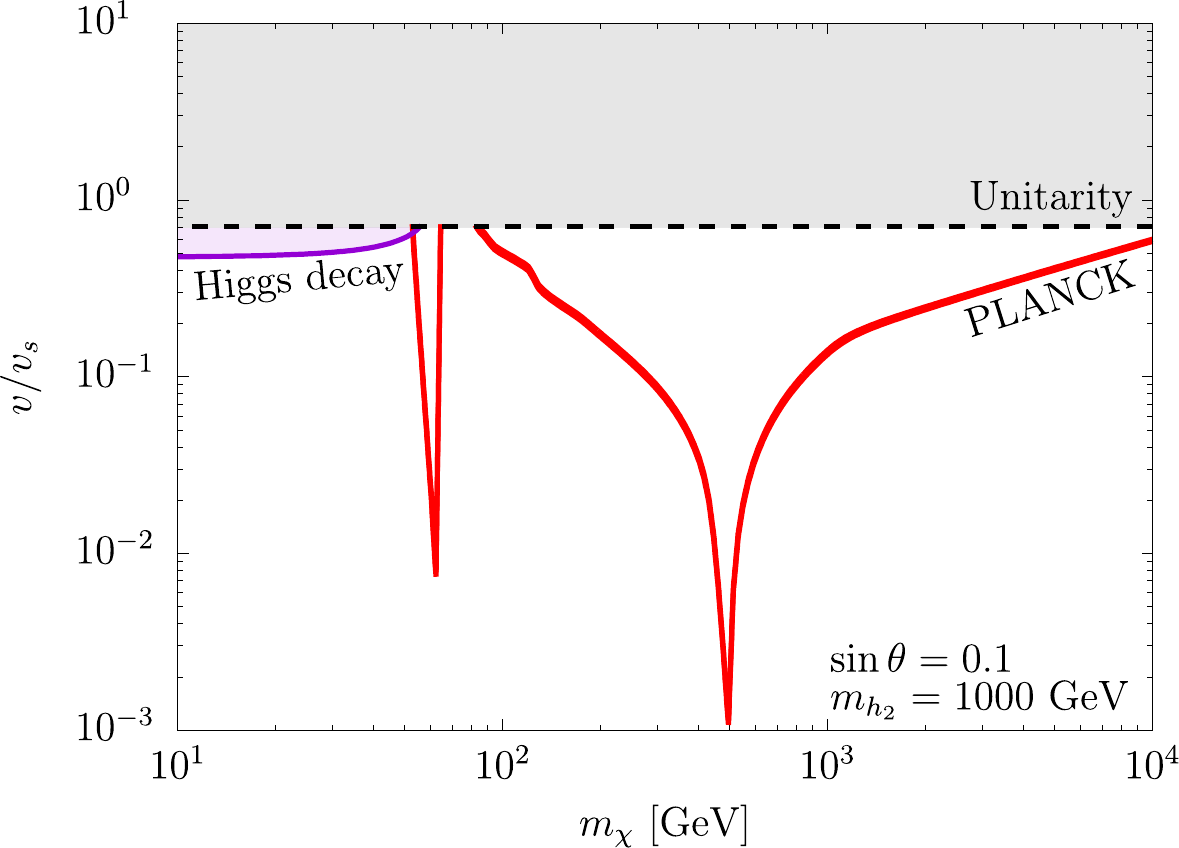}
 }
\caption{Allowed range of dark matter mass $m_\chi$ vs $v/v_s$. The red band corresponds to the thermal DM relic abundance consistent with the PLANCK measurements. The purple region
is excluded by the Higgs invisible decay constraint, while the perturbative unitarity bound is marked by the dashed line.
}
\label{parameter-space}
\end{figure}

The plots in Fig.~\ref{parameter-space} show the allowed parameter space in the plane ($m_\chi$,
$v/v_s$), where the mixing angle and the second Higgs mass are fixed to
be $\sin\theta=0.1$ and $m_{h_2}=300,1000$ GeV. The latter are consistent with the electroweak precision measurements and the Higgs data~\cite{Falkowski:2015iwa}.
The red curve corresponds to the correct relic DM abundance. It features the usual resonant annihilation dips at $m_{h_1}/2$ and $m_{h_2}/2$. The main DM annihilation channels are: $\chi\chi\to
b\overline{b},c\overline{c}$ for $m_{\chi} \lesssim m_{W}$; 
 $\chi\chi\to
W^+W^-,ZZ,h_1h_1 $ for $ m_{W} \lesssim m_{\chi} \lesssim m_{h_2}$;
$\chi\chi\to h_2h_2$ for $m_{h_2} \lesssim m_{\chi}$. These are not affected by the above described cancellation since the relevant momentum transfer is large, unlike that in the DM--nucleon scattering.
We see that the entire red band from $m_{\chi} \simeq m_{h_1}/2$ to 10~TeV is consistent with the other constraints.

The direct detection bounds are weak as expected from the loop suppression of the amplitude. For heavy dark matter, one can estimate an order of magnitude of the $\chi - N$ 
cross section by setting the loop functions to one,
\begin{equation}
\sigma_{\chi N} \sim {\sin^2\theta \over 64 \pi^5} \; { m_N^4 f_N^2 \over m_{h_1}^4 v^2 } \;
{ m_{h_2}^8 \over m_\chi^2 v_s^6 } \;,
\end{equation}
where $m_N$ is the nucleon mass and $f_N \sim 0.3$ parametrizes the Higgs--nucleon coupling. This gives $\sigma_{\chi N}$ in the ballpark of $10^{-49}$~cm$^{2}$ for $\sin\theta=0.1 , m_{h_2}=300 $ GeV and TeV dark matter mass. 
On the other hand, the best XENON1T limits are of order $10^{-46}$~cm$^{2}$.
For light dark matter, there is an additional suppression factor of order $m_\chi^4/m_{h_2}^4$ since in the limit $m_\chi \rightarrow 0 $ the U(1) symmetry is restored and the loop corrections vanish.
We thus find that the direct DM detection constraints are quite loose and in fact superseded by
the perturbative unitarity bound. 
The latter excludes
the upper parts of the plots since $v_s$ below or around the electroweak scale requires a large $\lambda_S$ to generate a given $m_{h_2}$.

All in all, the cancellation mechanism provides sufficient suppression of the direct detection amplitude such that the entire range of dark matter masses between 60 GeV and 10 TeV is allowed (depending on $m_{h_2}$ and $\sin\theta$).
Our main point is that although the cancellation affects the DM interaction with matter at zero momentum transfer, it does not apply at large momentum transfer relevant to DM annihilation processes.

Our dark matter candidate can potentially be detected at the LHC, for instance, via monojet events with missing energy. The analysis bears similarity to that of~\cite{Kim:2015hda}. 
In particular, one expects a substantial monojet rate when the ``heavy Higgs'' $h_2$ can decay into DM on-shell, i.e. $m_{h_2}>2 m_\chi$.
The kinematic reach, however, is likely to be limited to $m_{h_2}$ of order a few hundred GeV. 
A more detailed analysis will be presented elsewhere.

Another venue to probe the model at the LHC would be to study the Higgs couplings and search for a ``heavy Higgs'' $h_2$. The mixing of the Higgs with an SM singlet can be detected through universal reduction of the Higgs couplings, while $h_2$ would appear as a heavy Higgs-like resonance with reduced couplings to SM fields.

%=========================================================================
%=========================================================================
\section{U(1) breaking couplings as spurions} \label{spurion}
%=========================================================================
%=========================================================================

The presented scenario is expected to be a low energy limit of a more fundamental theory. 
Indeed, our model does not explain why the higher dimension terms such as $S^4$ are suppressed, why the odd powers of $S$ are absent and how an explicit symmetry breaking term can arise at all.
In the ultraviolet--complete model, the U(1) could be gauged and the symmetry breaking terms would result from spontaneous breaking. In what follows, let us leave aside the ``coincidence problem'' that $\mu_S \sim \mu_S'$ (akin to the $\mu$--problem of supersymmetry)
and focus on the hierarchy of the symmetry breaking couplings.

To illustrate our main point, consider a simplified model where the symmetry breaking terms are induced by a VEV of a single field $\Phi$ with charge $q_\Phi$, while $S$ has charge $q_S$.
If 
\be
n\equiv -2 q_S/ q_\Phi
\ee
is a positive odd number, the U(1) is broken down to a $Z_2$ subgroup such that interactions involving odd powers of $S$ are forbidden. 
For instance, an admissible choice would be $q_S=3$ and $q_\Phi=-2$.
Defining 
\be
\epsilon \equiv \frac{\langle \Phi \rangle}{\Lambda} \,,
\ee
with $\Lambda $ being some high energy scale associated with heavier states, and $\epsilon \ll 1$, U(1) invariance requires that the tree level couplings obey
\begin{equation}
\mu_S^{\prime 2} \sim \langle \Phi \rangle^2 \epsilon^{n-2} ~~,~~
\lhsp \sim \lspp \sim \epsilon^{n} ~~,~~ \lsp \sim \epsilon^{2n} \;.
\end{equation}

The magnitude of the couplings in Eq.~(26) is affected by loop corrections, in analogy with Sect.~4. For instance, the loop contribution to $\lambda_{HS}^\prime $ is proportional to $\mu^{\prime 2}_S$ times the loop factor. Thus, it is real for real $\mu^{\prime 2}_S$, although the tree level  $\lambda_{HS}^\prime $ in Eq.~(26) generally is not.

Clearly, the tree level $ \lhsp, \lspp $ and $\lsp$ can be made extremely small if the scale $\Lambda$
is very high. Let us estimate their lowest values. Given that we are interested in $\mu_S' \sim 100$ GeV, $ \epsilon^n \sim \mu_S^{\prime 2}/ \Lambda^2 \sim (100 \;{\rm GeV}/\Lambda)^2 \geq 10^{-32}$, where the lower bound is reached for $\Lambda$ close to the Planck scale. Thus $\lhsp, \lspp$ can be as small as $10^{-32}$ while $\lsp$ would be even smaller.

If the underlying dynamics conserve {\it CP}, the couplings are real and dark matter is stable 
due to the symmetry $S \rightarrow S^*$. However, one generally expects {\it CP} violation to be 
present even if suppressed. This introduces couplings linear in $\chi$ which 
 makes dark matter unstable albeit long lived. Let us estimate its longest possible lifetime taking into account only the most important factors: the $\epsilon$--suppression of the decay amplitude and the relevant DM scale of 
${\cal O}(100 \; {\rm GeV})$. One then has
\be
\tau_{\rm DM} \sim {8 \pi \over 100 \;{\rm GeV}}\; \epsilon^{-2n} \sim 10^{39}\; {\rm s} \;. \label{lifetime}
\ee
This is very much longer than the age of the Universe $\sim 10^{17}$s
and DM can be considered stable for all practical purposes.\footnote{Here, we have omitted another aspect of dark matter decay. Since both $\langle \Phi \rangle $
 and $\langle S \rangle $ break the U(1), dark matter has a tiny admixture of Im$\Phi$ as
the state orthogonal to the would-be Goldstone boson. The decay time of this component of dark matter is longer than that of Eq.~(\ref{lifetime}).}
This shows that the presence of U(1) breaking terms is not dangerous if the underlying dynamics takes place at a high scale.

%=========================================================================
%=========================================================================
\section{Summary}
%=========================================================================
%=========================================================================

We have presented a simple extension of the Standard Model with a complex scalar featuring softly broken U(1) symmetry. The imaginary part of this scalar plays the role of dark matter.
The resulting tree level DM--nucleon scattering amplitude exhibits a perfect cancellation between the light and heavy Higgs contributions at zero momentum transfer. This can be traced to the fact that U(1) is only broken by a mass term, which is justified by treating the U(1) breaking couplings as spurions.
The cancellation does not persist at loop level and a small direct DM detection rate is thus generated. Our numerical analysis shows that a broad range of WIMP dark matter mass,
roughly from 60 GeV to 10 TeV, is allowed in this model.

\vspace{10pt}
\noindent
{\bf Acknowledgements} 

\noindent
C.G. and O.L. acknowledge support from the Academy of Finland, project {\it The Higgs Boson and the Cosmos}. 
T.T. acknowledges support from JSPS Fellowships for Research Abroad. We acknowledge useful communication with A. Beniwal.

{}

\end{document}